\documentclass{llncs}

\usepackage{multicol}

\usepackage{amsmath}
\usepackage{amsfonts}
\usepackage{amssymb}

\usepackage{graphicx}
\usepackage{epic}
\usepackage{eepic}
\usepackage{epsfig,float}
\usepackage{color}
\usepackage{tabularx}

\renewcommand{\le}{\leqslant}
\renewcommand{\ge}{\geqslant}

\newcommand{\ol}{\overline}
\newcommand{\eps}{\varepsilon}
\newcommand{\emp}{\emptyset}

\newcommand{\Sig}{\Sigma}

\newcommand{\noin}{\noindent}

\newcommand{\bi}{\begin{itemize}}
\newcommand{\ei}{\end{itemize}}
\newcommand{\be}{\begin{enumerate}}
\newcommand{\ee}{\end{enumerate}}
\newcommand{\bd}{\begin{description}}
\newcommand{\ed}{\end{description}}
\newcommand{\bq}{\begin{quote}}
\newcommand{\eq}{\end{quote}}

\newcommand{\cD}{{\mathcal D}}

\newcommand{\cN}{{\mathcal N}}

\title{Quotient Complexity of Star-Free Languages
\thanks{This work was supported by the Natural Sciences and Engineering Research Council of Canada under grant No.~OGP0000871
}
}

\author{Janusz~Brzozowski and Bo Liu
 }

\authorrunning{Brzozowski, Liu}   

\institute{David R. Cheriton School of Computer Science, University of Waterloo, \\
Waterloo, ON, Canada N2L 3G1\\
\{ {\tt brzozo, b23liu} \}{\tt @uwaterloo.ca} 
}

\begin{document}

\maketitle
\today
\begin{abstract}
The quotient complexity, also known as state complexity, of a regular language is the number of distinct left quotients of the language.
The quotient complexity of an operation is the maximal quotient complexity of the language resulting from the operation, as a function of the quotient complexities of the operands.
The class of star-free languages is the smallest class  containing the  finite languages and closed under boolean operations and concatenation.
We prove that the tight bounds on the quotient complexities of union, intersection, difference,  symmetric difference, concatenation, and star  for star-free languages are the same as those for regular languages, with some small exceptions, whereas the bound for reversal is $2^n-1$.
\bigskip

\noin
\end{abstract}

\section{Introduction}

The class of regular languages can be defined as the smallest class containing the finite languages and closed under union, concatenation and star. Since regular languages are also closed under complementation, one can redefine them as the smallest class containing the finite languages and closed under boolean operations, concatenation and star. In this new formulation, a natural question is that of the \emph{generalized star height} of a regular language, which is the minimum number of nested stars required to define the language when boolean operations are allowed. 
It is not clear who first considered the problem of generalized star height, but McNaughton and Papert reported in their 1971 monograph~\cite{McPa71} that this problem had been open ``for many years''.
There exist regular languages of star height 0 and 1, but it is not even known whether there exists a language of star height 2. See 
{\small\tt http://liafa.jussieu.fr/\~{}jep/Problemes/starheight.html.}

We consider regular languages of star height 0, which are also called \emph{star-free}. 
In 1965, Sch\"utzenberger proved~\cite{Sch65} that a language is star-free if and only if its syntactic monoid is \emph{group-free}, that is, has only trivial subgroups. An equivalent condition is that the minimal deterministic automaton of a star-free language is \emph{permutation-free}, that is, has only trivial permutations. Another point of view is that these automata are \emph{counter-free}, since they cannot count modulo any integer greater than 1. They can, however, \emph{count to a threshold}, that is $1, 2, \ldots n-1, n$ \emph{or more}.
Such automata are called \emph{aperiodic,} and this is the term that we use.

The \emph{state complexity} of a regular language~\cite{Yu01} is the number of states in the minimal deterministic finite automaton accepting that language. We prefer  the equivalent concept of \emph{quotient complexity}~\cite{Brz10}, which is the number of distinct left quotients of the language, because quotient complexity has some advantages.
The \emph{quotient complexity of an operation} in a subclass of regular languages is the maximal quotient complexity of the language resulting from the operation, as a function of the quotient complexities of the operands when they range over all the languages in the subclass. 
The complexities of basic operations in the class of regular languages were studied by Maslov~\cite{Mas70} and Yu, Zhuang and Salomaa~\cite{YZS94}.

The complexities of operations were also considered in several subclasses of regular languages: 
unary~\cite{PiSh02,YZS94}, finite~\cite{CCSY01,Yu01}, ideal~\cite{BJL10}, closed~\cite{BJZ10}, prefix-free~\cite{HSW09},  suffix-free~\cite{HaSa09}, bifix-, factor-, and subword-free~\cite{BJS10},   and convex~\cite{Brz10a}. 
The complexity of operations can be significantly lower 
in a subclass of regular languages than in the general case. 
We prove that this is \emph{not} the case for star-free languages, which meet the bounds for regular languages, with  small exceptions.

It was shown in~\cite{BHK09} that the tight bound for converting an $n$-state aperiodic nondeterministic automaton to a deterministic one is $2^n$.

In Section~\ref{sec:terminology} we define our terminology and notation.
Boolean operations, concatenation, star,
and reversal  are studied in Sections~\ref{sec:boolean}--\ref{sec:reversal}, respectively.
Unary languages are treated in Section~\ref{sec:unary}, and
Section~\ref{sec:conclusions} concludes the paper.

\section{Terminology and Notation}
\label{sec:terminology}
If $\Sig$ is a finite non-empty alphabet, then $\Sig^*$ is the set of all  words over this alphabet,
with $\eps$ as the empty word. 
For $w\in\Sig^*$, $a\in\Sig$, let $|w|$ be the length of $w$, and  $|w|_a$, the number of $a$'s in $w$. A~language is any subset of $\Sig^*$.

We use the following set operations on languages:  {\em complement\/} ($\ol{L}=\Sig^*\setminus L$),  {\em union\/}  ($K\cup L$),  {\em intersection\/} ($K\cap L$),  {\em difference\/} ($K\setminus L$), and {\em symmetric difference\/} ($K\oplus L$). 
We also  use \emph{product}, also called  \emph{(con)catenation}  ($KL=\{w\in \Sig^*\mid w=uv, u\in K, v\in L\}$)  and   \emph{star} ($K^*=\bigcup_{i\ge 0}K^i$).
The reverse $w^R$ of a word $w\in\Sig^*$ is defined by: $\eps^R=\eps$, and $(wa)^R=aw^R$.
The \emph{reverse} of a language $L$ is 
$L^R=\{w^R\mid w\in L\}$.

\emph{Regular} languages are the smallest class of languages containing the finite languages and closed under boolean operations, product  and  star. 
\emph{Star-free} languages are the languages one can construct from finite languages using only boolean operations and concatenation.
Some examples of star-free languages are $\emp$, $\Sig^*=\ol{\emp}$, 
$b^*=\ol{\Sig^*a\Sig^*}= \ol {\ol{\emp}a\ol{\emp}}$ over $\Sig=\{a,b\}$,  and
$aa^*=\ol{\eps}$ over $\Sig=\{a\}$. We do not write such expressions for star-free  languages, but denote them as usual.

The \emph{(left) quotient} of a language $L$ by a word $w$ is defined as $L_w=\{x\in \Sig^* \mid wx \in L \}$. The number of distinct quotients of a language  is called its \emph{quotient complexity}  and is denoted by $\kappa(L)$. 
A~quotient $L_w$ is \emph{accepting} if $\eps\in L_w$; 
otherwise it is \emph{rejecting}.

A \emph{deterministic finite automaton (DFA)} is a quintuple $\cD=(Q,\Sigma,\delta,q_0,F)$, where $Q$ is a finite set of \emph{states,} $\Sigma$ is a finite \emph{alphabet,} $\delta:Q\times\Sigma\rightarrow Q$ is the \emph{transition function,} $q_0$ is the \emph{initial state,} and $F\subseteq Q$ is the set of \emph{final} or \emph{accepting states\/}. 
As usual, the transition function is extended to $Q\times\Sigma^*$. 
A DFA $\cD$ accepts $w\in\Sigma^*$ if ${\delta}(q_0,w)\in F$, and the \emph{language} accepted by $\cD$ is $L(\cD)$. 
The \emph{language of a state} $q$ of $\cD$ is the language $L_q$ accepted by the automaton $(Q,\Sigma,\delta,q,F)$. 
If the language of a state is empty, that state is \emph{empty}.

Let $L^\eps=\eps$ if $\eps\in L$, and $L^\eps=\emp$, otherwise.
The \emph{quotient automaton} of a regular language $L$ is 
$\cD=(Q, \Sig, \delta, q_0,F)$, where $Q=\{L_w\mid w\in\Sig^*\}$, $\delta(L_w,a)=L_{wa}$, 
$q_0=L_\eps=L$,  $F=\{L_w\mid L_w^\eps=\eps\}$, and $L_w^\eps=(L_w)^\eps$.
Since this  is the minimal DFA accepting $L$, the quotient complexity of $L$ is equal to the state complexity of $L$, and we call it simply  \emph{complexity}.

A \emph{transformation} of a set $S=\{1,\ldots,n\}$ into itself is a mapping 
$$
t=\left( \begin{array}{ccccc}
1 & 2 &   \cdots &  n-1 & n \\
i_1 & i_2 &   \cdots &  i_{n-1} & i_n
\end{array} \right ),
$$
where $i_k\in S$ for $1\le k\le n$. 
Each word in $\Sig^*$ performs a transformation of the set $Q$ of states of a DFA $\cD$.
A DFA is \emph{aperiodic} if no word performs a permutation, other than the identity permutation, of a subset of $Q$.
Since testing if a DFA is aperiodic is PSPACE-complete~\cite{ChHu91}, we  use a subclass of aperiodic automata. 
Without loss of generality, we assume that $Q=\{1,\ldots,n\}$.
A transformation is \emph{non-decreasing} if $j< k$ implies $i_j\le i_k$.
A~non-decreasing transformation cannot have a non-trivial permutation, and the composition of non-decreasing transformations is non-decreasing. 
Hence a DFA with non-decreasing input transformations is aperiodic.

A \emph{nondeterministic finite automaton  (NFA)} is defined as a quintuple
$\mathcal{N}=(Q,\Sigma,\eta,I,F),$
 where 
 $Q$, $\Sigma$, and $F$ are as in a DFA, 
 $\eta:Q\times\Sigma\rightarrow 2^{Q}$ is the \emph{transition function} and
 $I\subseteq Q$ is the \emph{set of initial states}. 
If $\eta$ also allows $\eps$, 
that is, $\eta:Q\times(\Sigma\cup\{\eps\})\rightarrow 2^{Q}$, 
we call $\cN$ an \emph{$\eps$-NFA}.

\section{Boolean Operations}
\label{sec:boolean}
We now consider the quotient complexity of  union, intersection, symmetric difference, and difference in the class of star-free languages. 
The upper bound for these four operations in the class of regular languages is $mn$~\cite{Brz10,Mas70,YZS94}.

\begin{theorem}
\label{thm:boolean}
For each of the operations  union, intersection, symmetric difference, and difference, there exist binary star-free languages $K$ and $L$ with quotient complexities $m\ge 1$ and $n\ge 1$, respectively, that meet the bound $mn$.
\end{theorem}

\begin{proof}

Let $\Sig=\{a,b\}$. We examine union first. 
For $m=1$, let $K=\emp$ and let $L$ be any binary star-free language with 
$\kappa(L)=n$.
Then $\kappa(K\cup L)=\kappa(L)=n=mn$.
Similarly, if $n=1$, let $L=\emp$ and let $K$ be any binary star-free language with 
$\kappa(K)=m$. Then $\kappa(K\cup L)=mn$.

For $m,n\ge 2$, let $K=(b^*a)^{m-2}b^*=\{w\in \Sig^*\mid |w|_a=m-2\}$, and 
$L=(a^*b)^{n-2}a^*=\{w\in\Sig^*\mid |w|_b=n-2\}$; then $\kappa(K)=m$ and $\kappa(L)=n$, and both $K$ and $L$ are star-free. 
The quotient automata of $K$ and $L$ are  in Fig.~\ref{fig:unionKL} for $m=4$ and $n=5$,
and  their direct product  for $K\cup L$, in Fig.~\ref{fig:union}.

 \begin{figure}[t]
\begin{center}
\setlength{\unitlength}{0.00039370in}
\begingroup\makeatletter\ifx\SetFigFont\undefined%
\gdef\SetFigFont#1#2#3#4#5{%
  \reset@font\fontsize{#1}{#2pt}%
  \fontfamily{#3}\fontseries{#4}\fontshape{#5}%
  \selectfont}%
\fi\endgroup%
{\renewcommand{\dashlinestretch}{30}
\begin{picture}(10794,1367)(0,-10)
\put(3286,257){\makebox(0,0)[lb]{\smash{{\SetFigFont{9}{10.8}{\rmdefault}{\mddefault}{\updefault}$3$}}}}
\put(3369.500,772.929){\arc{394.717}{2.4948}{6.9299}}
\blacken\thicklines
\path(3529.033,794.097)(3527.000,654.000)(3600.998,772.977)(3529.033,794.097)
\thinlines
\put(2206.500,750.929){\arc{394.717}{2.4948}{6.9299}}
\blacken\thicklines
\path(2366.033,772.097)(2364.000,632.000)(2437.998,750.977)(2366.033,772.097)
\thinlines
\put(1044.500,757.929){\arc{394.717}{2.4948}{6.9299}}
\blacken\thicklines
\path(1204.033,779.097)(1202.000,639.000)(1275.998,757.977)(1204.033,779.097)
\thinlines
\put(7225.500,739.929){\arc{394.717}{2.4948}{6.9299}}
\blacken\thicklines
\path(7385.033,761.097)(7383.000,621.000)(7456.998,739.977)(7385.033,761.097)
\thinlines
\put(8305.500,731.929){\arc{394.717}{2.4948}{6.9299}}
\blacken\thicklines
\path(8465.033,753.097)(8463.000,613.000)(8536.998,731.977)(8465.033,753.097)
\thinlines
\put(9385.500,731.929){\arc{394.717}{2.4948}{6.9299}}
\blacken\thicklines
\path(9545.033,753.097)(9543.000,613.000)(9616.998,731.977)(9545.033,753.097)
\thinlines
\put(10472.500,739.929){\arc{394.717}{2.4948}{6.9299}}
\blacken\thicklines
\path(10632.033,761.097)(10630.000,621.000)(10703.998,739.977)(10632.033,761.097)
\thinlines
\put(6145.500,724.929){\arc{394.717}{2.4948}{6.9299}}
\blacken\thicklines
\path(6305.033,746.097)(6303.000,606.000)(6376.998,724.977)(6305.033,746.097)
\thinlines
\put(1044,356){\ellipse{630}{630}}
\put(2215,343){\ellipse{630}{630}}
\put(3372,365){\ellipse{540}{540}}
\put(4537,354){\ellipse{630}{630}}
\put(3369,363){\ellipse{630}{630}}
\put(8312,329){\ellipse{630}{630}}
\put(7219,339){\ellipse{630}{630}}
\put(10471,325){\ellipse{630}{630}}
\put(9394,330){\ellipse{540}{540}}
\put(9391,329){\ellipse{630}{630}}
\put(6141,323){\ellipse{630}{630}}
\path(2537,347)(3047,347)
\blacken\thicklines
\path(2912.000,309.500)(3047.000,347.000)(2912.000,384.500)(2912.000,309.500)
\thinlines
\path(1374,354)(1884,354)
\blacken\thicklines
\path(1749.000,316.500)(1884.000,354.000)(1749.000,391.500)(1749.000,316.500)
\thinlines
\path(3692,354)(4202,354)
\blacken\thicklines
\path(4067.000,316.500)(4202.000,354.000)(4067.000,391.500)(4067.000,316.500)
\thinlines
\path(444,358)(714,358)
\blacken\thicklines
\path(579.000,320.500)(714.000,358.000)(579.000,395.500)(579.000,320.500)
\thinlines
\path(7542,336)(7992,336)
\blacken\thicklines
\path(7857.000,298.500)(7992.000,336.000)(7857.000,373.500)(7857.000,298.500)
\thinlines
\path(8629,343)(9079,343)
\blacken\thicklines
\path(8944.000,305.500)(9079.000,343.000)(8944.000,380.500)(8944.000,305.500)
\thinlines
\path(9702,336)(10152,336)
\blacken\thicklines
\path(10017.000,298.500)(10152.000,336.000)(10017.000,373.500)(10017.000,298.500)
\thinlines
\path(6447,343)(6897,343)
\blacken\thicklines
\path(6762.000,305.500)(6897.000,343.000)(6762.000,380.500)(6762.000,305.500)
\thinlines
\path(5530,336)(5800,336)
\blacken\thicklines
\path(5665.000,298.500)(5800.000,336.000)(5665.000,373.500)(5665.000,298.500)
\put(4277,1097){\makebox(0,0)[lb]{\smash{{\SetFigFont{9}{10.8}{\familydefault}{\mddefault}{\updefault}$a,b$}}}}
\put(985,1089){\makebox(0,0)[lb]{\smash{{\SetFigFont{9}{10.8}{\familydefault}{\mddefault}{\updefault}$b$}}}}
\put(2132,1103){\makebox(0,0)[lb]{\smash{{\SetFigFont{9}{10.8}{\familydefault}{\mddefault}{\updefault}$b$}}}}
\put(3264,1090){\makebox(0,0)[lb]{\smash{{\SetFigFont{9}{10.8}{\familydefault}{\mddefault}{\updefault}$b$}}}}
\put(3797,475){\makebox(0,0)[lb]{\smash{{\SetFigFont{9}{10.8}{\familydefault}{\mddefault}{\updefault}$a$}}}}
\put(2605,467){\makebox(0,0)[lb]{\smash{{\SetFigFont{9}{10.8}{\familydefault}{\mddefault}{\updefault}$a$}}}}
\put(1486,482){\makebox(0,0)[lb]{\smash{{\SetFigFont{9}{10.8}{\familydefault}{\mddefault}{\updefault}$a$}}}}
\put(2117,264){\makebox(0,0)[lb]{\smash{{\SetFigFont{9}{10.8}{\rmdefault}{\mddefault}{\updefault}$2$}}}}
\put(4449,249){\makebox(0,0)[lb]{\smash{{\SetFigFont{9}{10.8}{\rmdefault}{\mddefault}{\updefault}$4$}}}}
\put(976,272){\makebox(0,0)[lb]{\smash{{\SetFigFont{9}{10.8}{\rmdefault}{\mddefault}{\updefault}$1$}}}}
\put(15,262){\makebox(0,0)[lb]{\smash{{\SetFigFont{9}{10.8}{\rmdefault}{\mddefault}{\updefault}$K$}}}}
\put(6611,474){\makebox(0,0)[lb]{\smash{{\SetFigFont{9}{10.8}{\familydefault}{\mddefault}{\updefault}$b$}}}}
\put(5208,241){\makebox(0,0)[lb]{\smash{{\SetFigFont{9}{10.8}{\rmdefault}{\mddefault}{\updefault}$L$}}}}
\put(7130,256){\makebox(0,0)[lb]{\smash{{\SetFigFont{9}{10.8}{\rmdefault}{\mddefault}{\updefault}$2$}}}}
\put(8231,249){\makebox(0,0)[lb]{\smash{{\SetFigFont{9}{10.8}{\rmdefault}{\mddefault}{\updefault}$3$}}}}
\put(9311,249){\makebox(0,0)[lb]{\smash{{\SetFigFont{9}{10.8}{\rmdefault}{\mddefault}{\updefault}$4$}}}}
\put(10392,234){\makebox(0,0)[lb]{\smash{{\SetFigFont{9}{10.8}{\rmdefault}{\mddefault}{\updefault}$5$}}}}
\put(6072,249){\makebox(0,0)[lb]{\smash{{\SetFigFont{9}{10.8}{\rmdefault}{\mddefault}{\updefault}$1$}}}}
\put(7698,474){\makebox(0,0)[lb]{\smash{{\SetFigFont{9}{10.8}{\familydefault}{\mddefault}{\updefault}$b$}}}}
\put(8786,474){\makebox(0,0)[lb]{\smash{{\SetFigFont{9}{10.8}{\familydefault}{\mddefault}{\updefault}$b$}}}}
\put(9851,459){\makebox(0,0)[lb]{\smash{{\SetFigFont{9}{10.8}{\familydefault}{\mddefault}{\updefault}$b$}}}}
\put(6050,1059){\makebox(0,0)[lb]{\smash{{\SetFigFont{9}{10.8}{\familydefault}{\mddefault}{\updefault}$a$}}}}
\put(7122,1052){\makebox(0,0)[lb]{\smash{{\SetFigFont{9}{10.8}{\familydefault}{\mddefault}{\updefault}$a$}}}}
\put(8179,1059){\makebox(0,0)[lb]{\smash{{\SetFigFont{9}{10.8}{\familydefault}{\mddefault}{\updefault}$a$}}}}
\put(9260,1052){\makebox(0,0)[lb]{\smash{{\SetFigFont{9}{10.8}{\familydefault}{\mddefault}{\updefault}$a$}}}}
\put(10225,1066){\makebox(0,0)[lb]{\smash{{\SetFigFont{9}{10.8}{\familydefault}{\mddefault}{\updefault}$a,b$}}}}
\thinlines
\put(4539.500,757.929){\arc{394.717}{2.4948}{6.9299}}
\blacken\thicklines
\path(4699.033,779.097)(4697.000,639.000)(4770.998,757.977)(4699.033,779.097)
\end{picture}
}
\end{center}
\caption{Witnesses $K$ and $L$  for union with $m=4$ and $n=5$.} 
\label{fig:unionKL}
\end{figure}
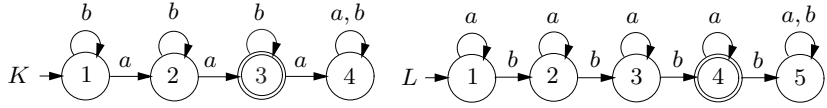

 \begin{figure}[t]
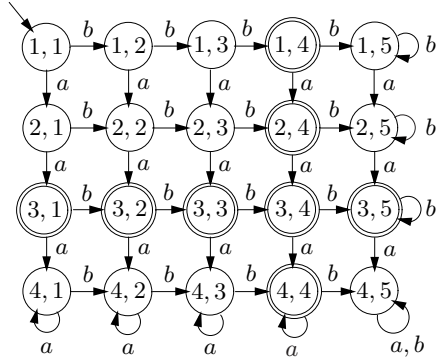

\begin{center}
\input union.eepic
\end{center}
\caption{Quotient automaton of $K\cup L$.} 
\label{fig:union}
\end{figure}

Let $M=K\cup L$, and consider the quotients of $M$ by the $mn$ words $a^ib^j$, $i=0,\ldots, m-1$, and $j=0,\ldots,n-1$;
these quotients $M_{a^ib^j}$ correspond to states $(i+1,j+1)$ in the direct-product automaton for $M$.
We begin with the rejecting quotients of $M$. 
First, $M_{a^{m-1}b^{n-1}}=\emp$, and all the other quotients are non-empty.
Next, if $i< m-2$ and $j<  n-2$ (rows 1 to $m-2$, columns 1 to $n-2$), then 
the  pair $(a^{m-2-i},b^{n-2-j})$ of non-empty words  belongs to $M_{a^ib^j}$ and to no other rejecting quotient.
If $i<  m-2$, then $M_{a^ib^{n-1}}$ (rows 1 to $m-2$, column $n$) contains $a^{m-2-i}$, but has no words from $b^*$.
If  $j< n-2$, then $M_{a^{m-1}b^j}$ (row $m$, columns 1 to $n-2$)  contains $b^{n-2-j}$, but has no words from $a^*$.
So all  rejecting  quotients are distinct.

Now turn to the accepting quotients.
For $i,k\le m-2$, quotient $M_{a^ib^{n-2}}$ (rows 1 to $m-1$, column $n-1$) contains $ba^{m-2-i}$, and this word is not contained in any other quotient $M_{a^kb^{n-2}}$ with $k\neq i$, and $M_{a^{m-1}b^{n-2}}$ has no words from $ba^*$. 
Thus all the quotients in column $n-1$ are distinct.
For $j,\ell \le n-2$,  $M_{a^{m-2}b^j}$ (row $m-1$, columns 1 to $n-1$) contains $ab^{n-2-j}$, and this word is not contained in any other quotient $M_{a^{m-2}b^\ell}$ with $\ell\neq j$, 
and $M_{a^{m-2}b^{n-1}}$ has no words from $ab^*$. 
Thus all the quotients in row $m-1$ are distinct.
Excluding $M_{a^{m-2}b^{n-2}}$, each quotient in  column $n-1$  contains $a$ but not $b$, each quotient in row $m-1$  contains $b$ but not $a$, and $M_{a^{m-2}b^{n-2}}$  contains both $a$ and $b$. Hence all accepting quotients are distinct, and
 our claim holds for union.
\goodbreak

For difference, we can use $\ol{K}$ and $L$, where $K$ and $L$ meet the bound $mn$ for union,
because
$\kappa(\ol{K}\setminus L)=\kappa (\ol{K}\cap \ol{L})=
\kappa(\ol{\ol{K}\cap \ol{L}})=\kappa(K\cup L)$.

For intersection, it was shown in~\cite{BJL10} that the languages 
$K=(b^*a)^{m-1}\Sig^*=\{w\in \Sig^*\mid |w|_a\ge m-1\}$ and
$L=(a^*b)^{n-1}\Sig^*=\{w\in \Sig^*\mid |w|_b\ge n-1\}$ meet the bound $mn$.
Since both languages are star-free, our claim holds for intersection.
These languages  also meet the bound $mn$ for symmetric difference~\cite{BJL10}.
\qed
\end{proof}

\section{Product}
\label{sec:product}

The tight bound for  product of  regular languages~\cite{Mas70,YZS94} is  $(m-1)2^n+2^{n-1}$.
We  show that this bound can be met by star-free languages, with some  exceptions. 

In subset constructions, we use the notation $S\stackrel{w}{\longrightarrow} T$ to mean that subset $S$ under input word $w$ moves to subset $T$.
\begin{theorem}
\label{thm:product}
There exist
quaternary star-free languages $K$ and $L$ with quotient complexities $m\ge 1$ and $n\ge 3$, respectively, such that  $\kappa(KL)=(m-1)2^n+2^{n-1}$.\end{theorem}
\begin{proof}
Let the quotient automaton for $K$ be $\cD_K=(Q_K,\Sig,\delta_K,q_0,F_K)$, where
$Q_K=\{q_1,q_2,\ldots,q_m\}$, $\Sig=\{a,b,c,d\}$, $q_0=q_1$, $F_K=\{q_{m}\}$, and 
\begin{eqnarray*}
\delta_K(q_i,a)&=&q_{i+1} \text{ for } i=1,\ldots,m-1,\quad  \delta_K(q_m,a)=q_m,\\
\delta_K(q_i,b)&=&q_{i-1} \text{ for } i=2,\ldots,m,\quad  \delta_K(q_1,b)=q_1,\\
\delta_K(q_i,c)&=&q_i \text{ for } i=1,\ldots,m,\\
\delta_K(q_i,d)&=&q_m \text{ for } i=1,\ldots,m.
\end{eqnarray*}

Next, let the quotient automaton for $L$ be $\cD_L=(Q_L,\Sig,\delta_L,p_0,F_L)$, where
$Q_L=\{1,2,\ldots,n\}$, $\Sig=\{a,b,c,d\}$, $p_0=1$, $F_L=\{n-1\}$, and 
\begin{eqnarray*}
\delta_L(i,c)&=&i+1 \text{ for } i=1,\ldots,n-1,\quad  \delta_L(n,c)=n,\\
\delta_L(i,d)&=&i-1 \text{ for } i=2,\ldots,n,\quad  \delta_L(1,d)=1,\\
\delta_L(i,a)&=&i+1 \text{ for } i=2,\ldots,n-1,\quad  \delta_L(1,a)=1,  \quad \delta_L(n,a)=n, \\
\delta_L(i,b)&=&i \text{ for } i=1,\ldots,n.
\end{eqnarray*}
The automaton $\cD_K$ for $m=4$ is shown in Fig.~\ref{fig:product}, where the transition labeled
$\eps$ should be ignored for now. 
The automaton $\cD_L$ for $n=5$ is also shown in  Fig.~\ref{fig:product}. 
If the transition labeled $\eps$ is taken into account and $q_4$ is made a rejecting state, then we have an $\eps$-NFA for $KL$.
Here the initial state is $q_1$, the set of accepting states is $\{4\}$, and the transitions are as shown.

\begin{figure}[t]
\begin{center}
\input product.eepic
\end{center}
\caption{$\eps$-NFA $\cN$ of $KL$.} 
\label{fig:product}
\end{figure}

For  $1\le s_k\le n-1$, $S=\{s_1,\ldots,s_k\}$, $s_1<s_2<\cdots <s_k$, $s_i\in Q_L$, and $0\le x\le n-s_k$,  denote $(s_1+x,\ldots,s_k+x)$ by $S_{+x}$.
Similarly, for  $2\le s_1\le n$,  and $0\le x\le s_1-1$,  denote $(s_1-x,\ldots,s_k-x)$ by $S_{-x}$.

We first show by induction on the size of $S$ that all $(m-1)2^{n-1}$ subsets of the form $\{q_i\}\cup S$, where $q_i\in Q_K$, $q_i\neq q_m$, and $S\subseteq Q_L\setminus \{1\}$, are reachable. 

When $S=\emp$, the set $\{q_i\}$ is reached by $a^{i-1}$, for $i=1,\ldots,m-1$.
Now suppose we want to reach $\{q_i\}\cup T$, where $i\neq m$, $T=\{s_0,s_1,\ldots,s_k\}$, 
$k\ge 0$, and $1< s_0<s_1<\cdots<s_k$. 
Let $S=\{s_1,\ldots,s_k\}$; by the induction assumption, $\{q_i\}\cup S$ is reachable. Then
$\{q_i\}\cup S\; \stackrel{d^{s_0-1}}{\longrightarrow} 
\{q_m,1\} \cup S_{-(s_0-1)}
\stackrel{b^{m-i}}{\longrightarrow} 
\{q_i,1\} \cup S_{-(s_0-1)}
\stackrel{c^{s_0-1}}{\longrightarrow}  
\{q_i\}\cup   \{s_0\} \cup S=
\{q_i\}\cup T\;$. 
Thus $\{q_i\}\cup T$ is also reachable.

Next, we prove that the $2^{n-1}$ subsets of the form $\{q_m,1\}  \cup S$, where $S$ is any subset of $Q_L\setminus \{1\}$, are reachable.
If $m=1$, then $\{q_1,1\}$ is the initial subset.
Let $S$ and $T$ be be as above.
Then $\{q_1,1\}\cup S \; \stackrel{d^{s_0-1}}{\longrightarrow} 
\{q_1,1\}\cup S_{-(s_0-1)} \stackrel{c}{\longrightarrow} 
\{q_1,1\}\cup \{2\}\cup  S_{-(s_0-2)} \stackrel{a^{s_0-2}}{\longrightarrow} 
\{q_1,1\}\cup \{s_0\}\cup  S=\{q_1,1\}\cup T
$.

If $m\ge 2$, there are two cases. If $2\not\in S$, then start with $\{q_1\}\cup S$, which has already been shown to be reachable. We then have
$\{q_1\}\cup S \stackrel{d}{\longrightarrow} 
\{q_m,1\} \cup S_{-1} \stackrel{a}{\longrightarrow} 
\{q_m,1\} \cup S.
$
If $2\in S$, then start with $\{q_1\}\cup S\setminus\{2\}$.
Now
$\{q_1\}\cup S\setminus\{2\} \stackrel{d}{\longrightarrow} 
\{q_m,1\}\cup (S\setminus\{2\})_{-1} \stackrel{c}{\longrightarrow}
\{q_m,1\}\cup \{2\} \cup (S\setminus\{2\}) 
=\{q_m,1\}\cup S$.

Finally, we show that the $(m-1)2^{n-1}$ subsets of the form 
$ \{q_i,1\}  \cup S$, 
where $i<m$, and $S\subseteq Q_L\setminus \{1\}$ are reachable. We have
$\{q_m,1\} \cup S 
\stackrel{b^{m-i}}{\longrightarrow} 
\{q_i,1\} \cup S$.

In summary, $(m-1)2^n+2^{n-1}$ different subsets are reachable.
We now prove that all these subsets are pairwise distinguishable.

For $1\le k\le n-1$, state $k$ of $Q_L$ accepts the word $w_k=c^{n-1-k}$, and state $n$ accepts the word $w_n=d$; moreover,  each of these words $w_h$ is accepted by only that one state $h$ of $Q_L$, and none of these words is accepted by state $q_i$, if $i\neq m$.
Hence, if $h$ is in $S\setminus T$ or in $T\setminus S$, then 
 $S$ and $T$  are distinguished by $w_h$.

First, let $1\le i \le j < m$, and consider $\{q_i\}\cup S$ and $\{q_j\}\cup T$, where  $S,T\subseteq Q_L$,  and $S$ and $T$ differ by state $h$.
Then $\{q_i\}\cup S$ and $\{q_j\}\cup T$ are distinguished by $w_h$.
Next, let $1\le i < j < m$ and  take $\{q_i\}\cup S$ and $\{q_j\}\cup S$, where  $S \subseteq Q_L$.
First apply $c$; then we reach $\{q_i\}\cup R$ and $\{q_j\}\cup R$, where
$1\not\in R$.
Then $\{q_j\}\cup R$ accepts $a^{m-j}c^{n-2}$, whereas $\{q_i\}\cup R$ rejects this word.

Second, suppose $S,T\subseteq Q_L\setminus \{1\}$ and  $S$ and $T$ differ by state $h$; then $\{q_m,1\}\cup S$ and $\{q_m,1\}\cup T$ are distinguished by $w_h$.

Third, consider $\{q_i\}\cup S$, where $S\subseteq Q_L$ and $\{q_m,1\}\cup T$, where $T\subseteq Q_L\setminus \{1\}$ and  $i<m$. 
Then  $c^{n-1}$ is accepted by $\{q_m,1\}\cup T$ but not by $ \{q_i\}\cup S$. 

Since all reachable sets are pairwise distinguishable,   the bound is met.
\qed
\end{proof}

\begin{corollary}
\label{cor:prod}
There exists a ternary star-free language $L$ with quotient complexity  $n\ge 1$, such that  $\kappa(\Sig^*L)=2^{n-1}$.
\end{corollary}
\begin{proof}
If $K=\Sig^*$,  the DFA $\cD_K$ has one  state, which is both initial and accepting.
Now  $b$ is not needed in the proofs of reachability and
distinguishability.
\qed
\end{proof}

A \emph{right (left) ideal}~\cite{BJL10} is a language $L$ satisfying $L=L\Sig^*$ ($L=\Sig^*L$). 
If $M=K\Sig^*$ ($M=\Sig^*K$), then $M$ is the right (left) ideal generated by $K$.
Corollary~\ref{cor:prod} shows that  the bound $2^{n-1}$ on the quotient complexity of the left ideal generated by a regular language can also be met by a star-free language.

If $n=1$ in Theorem~\ref{thm:product}, then either $KL=\emp$ and $\kappa(KL)=1$, or $KL=K\Sig^*$ is the right ideal generated by $K$. In the second case, it is known~\cite{YZS94} that $m$ is a tight bound for $\kappa(K\Sig^*)$, and that the language $a^{m-1}a^*$ is a witness~\cite{BJL10}. Since that witness is star-free, the general bound holds also for star-free languages.

The case $m\ge 2 $ and $n=2$ remains. 
For $m=n=2$, the best bound for product of  regular languages is 6, whereas it is 4 for star-free languages. This was verified with the \emph{GAP} package \emph{Automata}~\cite{GAP} by enumerating all products of 2-state aperiodic automata.

There are only three types of inputs possible for a 2-state aperiodic DFA:
the input that takes both states to state 1, the input that takes both states to state 2, and the identity input.
If 1 is the accepting state, then subsets $\{1\}$ and $\{1,2\}$ are not distinguishable.
Therefore a rejecting quotient  of $\cD_K$ can appear with only three subsets of quotients of $\cD_L$ in the DFA of $KL$ instead of $2^2=4$,  and an accepting quotient, only with one subset instead of two.
The complexity is maximized when there is only one accepting quotient of $K$. Hence $\kappa(KL)\le (m-1)3+1=3m-2$. 
If 2 is the accepting state, then  $\{2\}$ and $\{1,2\}$ are not distinguishable.
Hence $\kappa(KL)\le (m-1)3+2=3m-1$ in this case.

\begin{theorem}
\label{thm:m2}
There exist ternary star-free languages $K$ and $L$ with quotient complexities $m\ge 2$ and $2$, respectively, such that $\kappa(KL)=3m-2$.
\end{theorem}
\begin{proof}
Let $\cD_K(a,b,c)$ be the DFA in the proof of Theorem~\ref{thm:product} restricted to input alphabet $\{a,b,c\}$.
Let $\cD_L=(\{1,2\},\{a,b,c\},\delta,1,\{1\})$, where
\begin{eqnarray*}
\delta_L(i,a)&=&i \text{ for } i=1,2,\\
\delta_L(i,b)&=& 1 \text{ for } i=1,2,\\
\delta_L(i,c)&=&2 \text{ for } i=1,2.
\end{eqnarray*}
For $i\neq m$, subset  $\{q_i\}$ is reached by $a^{i-1}$,  
$\{q_i\}\cup \{1\}$, by $a^{m-1}b^{m-i}$, 
and $\{q_i\}\cup \{2\}$, by $a^{m-1}b^{m-i}c$. 
Finally, $\{q_m\}\cup \{1\}$ is reached by $a^{m-1}$.
This gives $3m-2$ subsets.

For $i\neq m$,   $\{q_i\}$ accepts no words from $b^*$, $\{q_i\}\cup \{1\}$ accepts $\eps$, and $\{q_i\}\cup \{2\}$ accepts $b$ but not $\eps$. 
Hence subsets $\{q_i\}\cup S$ and $\{q_i\}\cup T$ with $i,j\neq m$,
$S,T \in \{\emp, \{1\},\{2\} \}$, and $S\neq T$, are distinguishable. 
Next, $\{q_i\}\cup S$ and $\{q_j\}\cup S$ with $i<j<m$ are distinguished by $ca^{m-j}$.
Also, $\{q_i\}$ and $\{q_i\}\cup \{2\}$ are distinguished from $\{q_m\}\cup \{1\}$
by $\eps$, and $\{q_i\}\cup \{1\}$ from $\{q_m\}\cup \{1\}$ by $c$. Therefore all $3m-2$ subsets are distinguishable.
\qed
\end{proof}
We do not know whether the bound $3m-1$ can be reached. However, we have verified with \emph{GAP} that it cannot be reached if $m=2$.

\section{Star}
\label{sec:star}

The following DFA plays a key part in finding bounds on the quotient complexities of stars of star-free languages.
Let $n\ge 3$, and $\cD_n=\cD_n(a,b,c,d)=(Q,\{a,b,c,d\},\delta,1, \{n-1\})$, where 
$Q=\{1,2,\ldots, n\}$ and 
\begin{eqnarray*}
\delta(i,a)&=&{i+1} \text{ for } i=1,\ldots,n-1,\quad  \delta(n,a)=n,\\
\delta(i,b)&=&{i-1} \text{ for } i=2,\ldots,n,\quad  \delta(1,b)=1,\\
\delta(i,c)&=&{i-1} \text{ for } i=2,\ldots,n-1, \quad \delta(1,c)=1, \quad \delta(n,c)=n,\\
\delta(i,d)&=&n \text{ for } i=1,\ldots,n.
\end{eqnarray*}
Since all the inputs perform non-decreasing transformations, $\cD_n$ is 
aperiodic. 

\begin{figure}[h]
\begin{center}
\setlength{\unitlength}{0.00039370in}
\begingroup\makeatletter\ifx\SetFigFont\undefined%
\gdef\SetFigFont#1#2#3#4#5{%
  \reset@font\fontsize{#1}{#2pt}%
  \fontfamily{#3}\fontseries{#4}\fontshape{#5}%
  \selectfont}%
\fi\endgroup%
{\renewcommand{\dashlinestretch}{30}
\begin{picture}(8975,2206)(0,-10)
\put(4332,1871){\makebox(0,0)[lb]{\smash{{\SetFigFont{9}{10.8}{\familydefault}{\mddefault}{\updefault}$\eps$}}}}
\put(1762.780,1308.139){\arc{331.829}{2.2779}{7.3005}}
\blacken\thicklines
\path(1893.711,1300.119)(1850.000,1167.000)(1956.106,1258.503)(1893.711,1300.119)
\thinlines
\put(2848,872){\ellipse{630}{630}}
\put(7218,875){\ellipse{630}{630}}
\put(7220,874){\ellipse{702}{702}}
\put(3918,874){\ellipse{630}{630}}
\put(6078,879){\ellipse{630}{630}}
\put(8295,818){\ellipse{630}{630}}
\put(642,874){\ellipse{702}{702}}
\put(641,872){\ellipse{630}{630}}
\put(5010,884){\ellipse{630}{630}}
\put(1754,874){\ellipse{630}{630}}
\path(1985,1099)(2615,1099)
\blacken\thicklines
\path(2480.000,1061.500)(2615.000,1099.000)(2480.000,1136.500)(2480.000,1061.500)
\thinlines
\path(2615,649)(1985,649)
\blacken\thicklines
\path(2120.000,686.500)(1985.000,649.000)(2120.000,611.500)(2120.000,686.500)
\thinlines
\path(3065,1099)(3695,1099)
\blacken\thicklines
\path(3560.000,1061.500)(3695.000,1099.000)(3560.000,1136.500)(3560.000,1061.500)
\thinlines
\path(3695,649)(3065,649)
\blacken\thicklines
\path(3200.000,686.500)(3065.000,649.000)(3200.000,611.500)(3200.000,686.500)
\thinlines
\path(4145,1099)(4775,1099)
\blacken\thicklines
\path(4640.000,1061.500)(4775.000,1099.000)(4640.000,1136.500)(4640.000,1061.500)
\thinlines
\path(5225,1099)(5855,1099)
\blacken\thicklines
\path(5720.000,1061.500)(5855.000,1099.000)(5720.000,1136.500)(5720.000,1061.500)
\thinlines
\path(6305,1099)(6935,1099)
\blacken\thicklines
\path(6800.000,1061.500)(6935.000,1099.000)(6800.000,1136.500)(6800.000,1061.500)
\thinlines
\path(7475,1099)(8105,1099)
\blacken\thicklines
\path(7970.000,1061.500)(8105.000,1099.000)(7970.000,1136.500)(7970.000,1061.500)
\thinlines
\path(4775,649)(4145,649)
\blacken\thicklines
\path(4280.000,686.500)(4145.000,649.000)(4280.000,611.500)(4280.000,686.500)
\thinlines
\path(5855,649)(5225,649)
\blacken\thicklines
\path(5360.000,686.500)(5225.000,649.000)(5360.000,611.500)(5360.000,686.500)
\thinlines
\path(6935,649)(6305,649)
\blacken\thicklines
\path(6440.000,686.500)(6305.000,649.000)(6440.000,611.500)(6440.000,686.500)
\thinlines
\path(12,882)(282,882)
\blacken\path(162.000,852.000)(282.000,882.000)(162.000,912.000)(162.000,852.000)
\path(987,874)(1437,874)
\blacken\thicklines
\path(1302.000,836.500)(1437.000,874.000)(1302.000,911.500)(1302.000,836.500)
\thinlines
\path(8015,657)(7497,657)
\blacken\thicklines
\path(7632.000,694.500)(7497.000,657.000)(7632.000,619.500)(7632.000,694.500)
\thinlines
\path(1985,619)(1992,619)
\path(1992,619)(1985,619)
\path(755,536)(756,535)(758,533)
	(761,529)(767,523)(774,514)
	(784,503)(797,489)(813,472)
	(830,453)(851,432)(874,409)
	(899,384)(925,359)(954,332)
	(984,305)(1016,278)(1049,252)
	(1083,225)(1119,200)(1156,175)
	(1195,151)(1236,128)(1280,107)
	(1325,87)(1373,69)(1424,53)
	(1478,39)(1535,27)(1595,19)
	(1658,14)(1722,12)(1783,14)
	(1843,20)(1902,28)(1959,39)
	(2014,53)(2067,68)(2117,85)
	(2166,104)(2213,124)(2258,145)
	(2301,167)(2343,191)(2384,215)
	(2424,240)(2463,265)(2500,291)
	(2537,317)(2571,342)(2605,368)
	(2637,393)(2666,416)(2694,439)
	(2719,460)(2742,479)(2762,496)
	(2779,510)(2793,523)(2804,532)
	(2812,540)(2825,551)
\blacken\path(2752.772,450.585)(2825.000,551.000)(2714.015,496.389)(2752.772,450.585)
\path(7205,1242)(7204,1242)(7203,1243)
	(7200,1245)(7196,1248)(7190,1252)
	(7182,1258)(7171,1265)(7158,1274)
	(7142,1284)(7123,1297)(7102,1311)
	(7077,1328)(7049,1346)(7019,1365)
	(6985,1387)(6949,1410)(6910,1434)
	(6869,1460)(6825,1486)(6779,1514)
	(6731,1542)(6680,1572)(6629,1601)
	(6575,1631)(6520,1661)(6464,1691)
	(6406,1721)(6347,1751)(6286,1781)
	(6225,1810)(6162,1838)(6097,1866)
	(6031,1894)(5964,1920)(5895,1946)
	(5825,1971)(5753,1995)(5679,2018)
	(5602,2040)(5524,2061)(5443,2081)
	(5360,2099)(5274,2115)(5186,2131)
	(5095,2144)(5002,2156)(4906,2165)
	(4808,2172)(4708,2177)(4607,2179)
	(4505,2179)(4403,2176)(4303,2170)
	(4204,2161)(4107,2150)(4012,2137)
	(3920,2122)(3831,2105)(3744,2087)
	(3660,2067)(3579,2046)(3501,2024)
	(3425,2000)(3351,1975)(3279,1949)
	(3210,1923)(3142,1895)(3076,1867)
	(3012,1838)(2949,1808)(2888,1778)
	(2828,1747)(2770,1715)(2712,1684)
	(2657,1652)(2602,1620)(2549,1588)
	(2497,1556)(2447,1524)(2398,1492)
	(2352,1461)(2307,1431)(2264,1402)
	(2223,1374)(2184,1347)(2148,1321)
	(2114,1297)(2083,1274)(2054,1253)
	(2029,1234)(2006,1217)(1985,1202)
	(1968,1189)(1953,1177)(1941,1168)
	(1931,1161)(1924,1155)(1918,1150)(1910,1144)
\blacken\thicklines
\path(1995.500,1255.000)(1910.000,1144.000)(2040.500,1195.000)(1995.500,1255.000)
\put(2772,807){\makebox(0,0)[lb]{\smash{{\SetFigFont{9}{10.8}{\rmdefault}{\mddefault}{\updefault}$2$}}}}
\put(3844,799){\makebox(0,0)[lb]{\smash{{\SetFigFont{9}{10.8}{\rmdefault}{\mddefault}{\updefault}$3$}}}}
\put(4925,815){\makebox(0,0)[lb]{\smash{{\SetFigFont{9}{10.8}{\rmdefault}{\mddefault}{\updefault}$4$}}}}
\put(5996,800){\makebox(0,0)[lb]{\smash{{\SetFigFont{9}{10.8}{\rmdefault}{\mddefault}{\updefault}$5$}}}}
\put(7136,793){\makebox(0,0)[lb]{\smash{{\SetFigFont{9}{10.8}{\rmdefault}{\mddefault}{\updefault}$6$}}}}
\put(2255,1181){\makebox(0,0)[lb]{\smash{{\SetFigFont{9}{10.8}{\familydefault}{\mddefault}{\updefault}$a$}}}}
\put(3223,363){\makebox(0,0)[lb]{\smash{{\SetFigFont{9}{10.8}{\familydefault}{\mddefault}{\updefault}$b,c$}}}}
\put(2067,373){\makebox(0,0)[lb]{\smash{{\SetFigFont{9}{10.8}{\familydefault}{\mddefault}{\updefault}$b,c$}}}}
\put(987,1038){\makebox(0,0)[lb]{\smash{{\SetFigFont{9}{10.8}{\familydefault}{\mddefault}{\updefault}$b,c$}}}}
\put(3313,1204){\makebox(0,0)[lb]{\smash{{\SetFigFont{9}{10.8}{\familydefault}{\mddefault}{\updefault}$a$}}}}
\put(4385,1211){\makebox(0,0)[lb]{\smash{{\SetFigFont{9}{10.8}{\familydefault}{\mddefault}{\updefault}$a$}}}}
\put(5450,1219){\makebox(0,0)[lb]{\smash{{\SetFigFont{9}{10.8}{\familydefault}{\mddefault}{\updefault}$a$}}}}
\put(6545,1219){\makebox(0,0)[lb]{\smash{{\SetFigFont{9}{10.8}{\familydefault}{\mddefault}{\updefault}$a$}}}}
\put(7565,1204){\makebox(0,0)[lb]{\smash{{\SetFigFont{9}{10.8}{\familydefault}{\mddefault}{\updefault}$a$}}}}
\put(4295,363){\makebox(0,0)[lb]{\smash{{\SetFigFont{9}{10.8}{\familydefault}{\mddefault}{\updefault}$b,c$}}}}
\put(5374,355){\makebox(0,0)[lb]{\smash{{\SetFigFont{9}{10.8}{\familydefault}{\mddefault}{\updefault}$b,c$}}}}
\put(6461,350){\makebox(0,0)[lb]{\smash{{\SetFigFont{9}{10.8}{\familydefault}{\mddefault}{\updefault}$b,c$}}}}
\put(7623,350){\makebox(0,0)[lb]{\smash{{\SetFigFont{9}{10.8}{\familydefault}{\mddefault}{\updefault}$b$}}}}
\put(8960,777){\makebox(0,0)[lb]{\smash{{\SetFigFont{9}{10.8}{\familydefault}{\mddefault}{\updefault}$a,c$}}}}
\put(1451,124){\makebox(0,0)[lb]{\smash{{\SetFigFont{9}{10.8}{\familydefault}{\mddefault}{\updefault}$a$}}}}
\put(559,785){\makebox(0,0)[lb]{\smash{{\SetFigFont{9}{10.8}{\rmdefault}{\mddefault}{\updefault}$0$}}}}
\put(8201,770){\makebox(0,0)[lb]{\smash{{\SetFigFont{9}{10.8}{\rmdefault}{\mddefault}{\updefault}$7$}}}}
\put(1684,792){\makebox(0,0)[lb]{\smash{{\SetFigFont{9}{10.8}{\rmdefault}{\mddefault}{\updefault}$1$}}}}
\put(1531,1594){\makebox(0,0)[lb]{\smash{{\SetFigFont{9}{10.8}{\familydefault}{\mddefault}{\updefault}$b,c$}}}}
\thinlines
\put(8740.333,822.000){\arc{333.333}{3.7851}{8.7813}}
\blacken\thicklines
\path(8743.529,690.520)(8607.000,722.000)(8707.733,624.614)(8743.529,690.520)
\end{picture}
}
\end{center}
\caption{$\eps$-NFA $\cN$ of $L^*$,  $\kappa(L)=7$. Transitions under $d$ (not shown) are all to state 7.} 
\label{fig:star}
\end{figure}
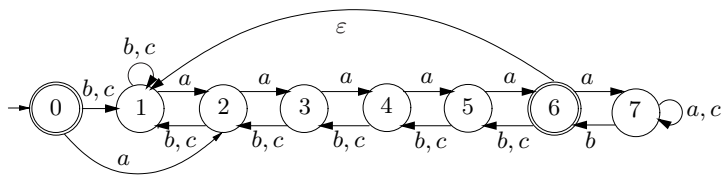

In Fig.~\ref{fig:star}, if we ignore state 0 and its outgoing transitions, and also the $\eps$ transition, then the figure
 shows the automaton $\cD_7(a,b,c,d)$. With state 0 and the $\eps$ transition it
depicts the $\eps$-NFA of $L^*$.

We first study  $\cD_n(a,b)$, the restriction of $\cD_n(a,b,c,d)$ to the alphabet $\{a,b\}$.
\begin{lemma}
\label{lem:star}
If $n\ge 3$, and $L$ is the  star-free language accepted by $\cD_n(a,b)$, then 
$\kappa(L^*)=2^{n-1}+2^{n-3}-1$. 
\end{lemma}
\begin{proof}
Consider the subsets of $\{0\}\cup Q$ in the subset construction of the DFA for $L^*$. Since 0 can only appear in $\{0\}$, the remaining reachable subsets are subsets of $Q$. 
The empty subset cannot be reached because there is a transition from each state under every letter. Since state $n-1$ cannot occur without state 1, we eliminate $2^{n-2}$ subsets.
Because state $n-1$ always appears with state $1$, and state $n$ can only be reached from state $n-1$ by $a$, the subset $\{n\}$ first appears with state 2, and afterwards, always with a state from $\{1,\ldots,n-1\}$; hence $\{n\}$ cannot be reached.
Also,   $1$ and $n$ cannot appear together without $n-1$, because  $n$ cannot be reached by $b$, and  1 cannot be reached by~$a$ without including $n-1$. This eliminates another $2^{n-3}$ subsets.
So $1+2^{n-2}+1+ 2^{n-3}$ subsets are unreachable, and $\kappa(L^*)\le 2^{n}+1- (2^{n-2}+2^{n-3}+2)=2^{n-1}+2^{n-3}-1$.

Now turn to the reachable subsets, and note that subsets $\{0\}$ and $\{1\}$ are reached by $\eps$ and $b$, respectively.

First, let $\mathbb{P}=\{S\subseteq \{2,\ldots,n-2\}\mid S\neq\emp\}$. 
 All singleton sets $\{i\}\in \mathbb{P} $ are reached by $a^{i-1}$ from $\{1\}$. 
 Now let  $S=\{s_1,\ldots,s_k\}$,
$T=\{s_0,s_1,\ldots,s_k\}$, where $0< k$,\;  $1< s_0< s_1<\cdots<s_k<n-1$, and 
$h=n-1-s_k$; then
$S \stackrel{a^{h}}{\rightarrow} \{1\}\cup S_{+h}
 \stackrel{b^{h}}{\rightarrow} \{1\}\cup S\;
 \stackrel{b^{s_0-1}}{\rightarrow} \{1\} \cup S_{-(s_0-1)}
\stackrel{a^{s_0-1}}{\rightarrow} \{s_0\} \cup S.
$
Thus any $T\in \mathbb{P}$ can be reached from a smaller $S\in \mathbb{P}$, and so all subsets in $\mathbb{P}$ are reachable.

Second, let $\mathbb{Q}=\{ \{1\}\cup S\mid S\in \mathbb{P} \}$; then 
$S \stackrel{a^hb^{h}}{\rightarrow} \{1\}\cup S$, as above,
and all  subsets in $\mathbb{Q}$ are reachable.

Third, let $\mathbb{R}=\{\{1,n-1\}\cup S \mid S=\emp  \text{ or } S\in \mathbb{P}\}$. 
If $S=\emp$, then $ \{1,n-1\}$ is reachable from $\{1\}$ by $a^{n-2}$. 
 Now suppose $S\in \mathbb{P}$ is not empty.
 If $i\in S$, then $\{i\}\stackrel{a^{n-1-i}}{\rightarrow} \{1,n-1\} \stackrel{a^{i-1}}{\rightarrow} \{i,n\}$. 
 So  $S\stackrel{a^{n-2}}{\rightarrow}  \{n\}\cup S$.
Now, if $s_k=n-2$, then 
$\{n\}\cup S \stackrel{a}{\rightarrow}    \{1, n-1,n\}\cup S_{+1} \stackrel{b}{\rightarrow}  \{1, n-1\}\cup S.$
If $s_k<n-2$, then 
$\{n\}\cup S \stackrel{a}{\rightarrow} \{n\}\cup S_{+1} 
 \stackrel{b}{\rightarrow}  \{1, n-1\}\cup S.$
In either case, 
$
S\stackrel{a^{n-1}b}{\rightarrow} \{1,n-1\}\cup S , 
$
and all $2^{n-3}$ subsets in $\mathbb{R}$ are reachable.

Fourth, let $\mathbb{S}= \{\{n\}\cup T \mid T\in \mathbb{P}\cup  \mathbb{R}\} $. We have shown that 
$S\stackrel{a^{n-2}}{\rightarrow} \{n\}\cup S$, if $S\in  \mathbb{P}$.
Since also 
$\{1,n-1\}\stackrel{a^{n-2}}{\rightarrow} \{1,n-1,n\}$, we have
$\{1,n-1\}\cup S \stackrel{a^{n-2}}{\rightarrow} \{1,n-1,n\}\cup S$.
Hence all $2^{n-2}-1$ subsets $\{n\}\cup T$ in $\mathbb{S}$ are reachable.

Altogether, $2^{n-1}+2^{n-3}-1$ subsets are reachable.
It remains to be shown that all the reachable subsets are pairwise distinguishable. 
State 0 does not accept $ab$, while $n-1$ accepts it. 
Each state 
$i$ with $1\le i\le n-2$ accepts $a^{n-1-i}$ and each of these words is  accepted by only that one state, and $n$ accepts $b$. So any two subsets $S$ and $T\neq S$ are distinguishable.
\qed
\end{proof}

\begin{theorem}
\label{thm:stars}
For $n\ge 2$ there exists a quaternary star-free language $L$ with $\kappa(L)=n$ such that $\kappa(L^*)=2^{n-1}+2^{n-2}$.
For $n=1$,  the tight upper bound is~2. 
\end{theorem}
\begin{proof}
For $n=1$, there are only two languages, $\emp$ and $\Sig^*$, and both are star-free. We have $\kappa(\emp^*)=2$, and $\kappa((\Sig^*)^*)=1$. For $n=2$, there are two star-free unary languages, $\eps$ and $aa^*$, and the bound cannot be met if $|\Sig|=1$.  If $\Sig=\{a,b\}$, then $b^*a\Sig^*$ meets the bound $3$.
For $n=3$, we analyzed all 3-state aperiodic automata using \emph{GAP}.
The bound 6 is met by $\cD_{3}(a,b,c,d)$ defined above, and  bounds 5 and 4 are met by $\cD_3(a,b,c)$  and $\cD_3(a,b)$, respectively.
These bounds cannot be improved.

We now turn to  the general case. 
We will show that the following sets of states are reachable in  the nondeterministic automaton $\cN$ (see Fig.~\ref{fig:star}) from the initial state 0: the set $\{0\}$, all subsets of $Q$ containing $\{1,n-1\}$, and all  
 non-empty subsets of $Q\setminus (n-1)$.
By Lemma~\ref{lem:star}, we can reach 
all these subsets by words in $\{a,b\}^*$, except 
$\{n\}$ and the subsets of $Q\setminus (n-1)$ containing $\{1,n\}$.

We have
$\{1,n-1\} \stackrel{a}{\rightarrow} \{2,n\} \stackrel{c}{\rightarrow} \{1,n\}$; hence $\{1,n\}$ is reachable.
Now consider $\{n\}\cup S$, where $S=\{s_1,s_2,\ldots, s_{k}\} \in \mathbb{P}$.
Let $h= n-1-s_{k}$; then using $a^h$ we move to 
 $\{1,n\}\cup S_{+h}$,  and by $c^h$ we reach
$T=\{1, n\}\cup S$. Since $\{n\}\cup S$ is reachable by Lemma~\ref{lem:star}, $T$ is also reachable. Thus we can reach all the subsets of $Q\setminus (n-1)$ containing $\{1,n\}$ by words in $\{a,b,c\}^*$.
The only set missing now is $\{n\}$, and it is reached by $d$.

In Lemma~\ref{lem:star}, we have already shown that any two subsets $S,T\subseteq Q$ such that $T\neq S$ are distinguishable by words in $\{a,b\}^*$.
\qed
\end{proof}

Table~\ref{tab:StarSummary} summarizes our results for the quotient complexity of $L^*$ in case $L$ is star-free. For unary languages, see Section~\ref{sec:unary}. The figures in boldface type are known to be tight upper bounds. 
For $n=4$, we analyzed all 4-state automata with non-decreasing input transformations.
Automata $\cD_{4}(a,b,c,d)$,  $\cD_{4}(a,b,c)$, and $\cD_{4}(a,b)$ meet the bounds 12,
11, and 9, respectively.
The bounds 11 and 9 cannot be improved in the class of automata with non-decreasing input transformations.
For the rest, the bounds for $|\Sig|=3$ and $|\Sig|=2$ are met by 
$\cD_{n}(a,b,c)$, and $\cD_{n}(a,b)$, respectively.

\setlength{\extrarowheight}{2pt}
\begin{table}[ht]
\caption{Quotient  complexities for stars of star-free languages.}
\label{tab:StarSummary}
\begin{center}
$
\begin{array}{| c ||c|c| c| c|c|c|c|c|c|c|}    
\hline
\ \ n \ \ &\ \ 1 \ \ &\ \ 2 \ \ &\ \ 3 \  \ & \  \  4 \ \ 
& \ \ 5 \ \  &\ \ 6 \ \  & \ \ 7 \ \ &\  \ 8 \  \ & \cdots 
&\  n \  
  \\
\hline  \hline
  |\Sig|=1
&\bf  2 & \bf 2  & \bf 3 & \bf 4	& \bf 5 & \bf   7	&\bf 13  &\bf 21   & \cdots &\ \bf n^2-7n+13 \ \\
\hline
|\Sig|=2 
& - &\bf 3  & \bf 4  &   9	&  19 & 39  & 79  &159   & \cdots & 2^{n-1}+2^{n-3}-1\\
\hline
|\Sig|=3
& - & -& \bf 5 &  11	& 23 & 47 & 95  &191   &\cdots  	&2^{n-1}+2^{n-2}-1  \\
\hline
|\Sig|=4
& - &-  & \bf 6 &\bf 12	& \bf 24 &  \bf 48  &\bf 96  &\bf 192  & \cdots	&\bf 2^{n-1}+2^{n-2} \\
\hline
\end{array}
$
\end{center}
\end{table}

\section{Reversal}
\label{sec:reversal}
For regular binary languages, the tight bound for reversal~\cite{Lei81}  is $2^n$. For star-free languages the bound  $2^n-1$ can be met, but with $|\Sig|=n-1$ letters.

\begin{theorem}
\label{thm:reversal}
For each $n\ge 1$ there exists a star-free language $L$ with quotient complexity $n$ such that $\kappa(L^R)=2^n-1$.
For $n=1$, the bound is met if $|\Sig|\ge 1$,  for $n=2$, if $|\Sig|\ge 2$, and for $n\ge 3$, if $|\Sig|\ge n-1$.
\end{theorem}
\begin{proof}
For $n=1$ and $\Sig=\{a\}$, $a^*$ is a witness. 
For $n=2$ and $\Sig=\{a,b\}$, $\Sig^*a$ is a witness. 
We have verified using \emph{GAP} that all star-free languages $L$ with $n=2$ 
satisfy $\kappa(L^R)\le 3$; hence this bound cannot be increased.

Now let $n\ge 3$, and let 
$\cD_n=(Q,\Sig,\delta,1, E),$
 where 
$Q=\{1,2,\ldots, n\}$, $\Sig=\{a,b,c_3,\ldots,c_{n-1}\}$,  $E=  \{i\in Q\mid i \text{ is even}\}$, and 
\begin{eqnarray*}
\delta(i,a)&=&{i+1} \text{ for } i=1,\ldots,n-1,\quad  \delta(n,a)=n,\\
\delta(i,b)&=&i-1 \text{ for } i=2,\ldots,n,\quad  \delta(1,b)=1,\\
\delta(i,c_j)&=&i \text{ for } i\not=j, \quad \delta(j,c_j)=j-1 \quad \text{ for } j=3,\ldots,n-1.
\end{eqnarray*}
Since all the inputs perform non-decreasing transformations, $\cD_n$ is 
aperiodic. 
Figure~\ref{fig:reversalodd} shows the NFA $\cN$ which is the reverse of  DFA $\cD_7$.

\begin{figure}[tbh]
\begin{center}
\setlength{\unitlength}{0.00039370in}
\begingroup\makeatletter\ifx\SetFigFont\undefined%
\gdef\SetFigFont#1#2#3#4#5{%
  \reset@font\fontsize{#1}{#2pt}%
  \fontfamily{#3}\fontseries{#4}\fontshape{#5}%
  \selectfont}%
\fi\endgroup%
{\renewcommand{\dashlinestretch}{30}
\begin{picture}(11618,2051)(0,-10)
\put(10337,350){\makebox(0,0)[lb]{\smash{{\SetFigFont{9}{10.8}{\familydefault}{\mddefault}{\updefault}$b$}}}}
\put(11296.500,1218.929){\arc{394.717}{2.4948}{6.9299}}
\blacken\thicklines
\path(11456.033,1240.097)(11454.000,1100.000)(11527.998,1218.977)(11456.033,1240.097)
\thinlines
\put(7869.500,1233.929){\arc{394.717}{2.4948}{6.9299}}
\blacken\thicklines
\path(8029.033,1255.097)(8027.000,1115.000)(8100.998,1233.977)(8029.033,1255.097)
\thinlines
\put(6181.500,1226.929){\arc{394.717}{2.4948}{6.9299}}
\blacken\thicklines
\path(6341.033,1248.097)(6339.000,1108.000)(6412.998,1226.977)(6341.033,1248.097)
\thinlines
\put(4449.500,1233.929){\arc{394.717}{2.4948}{6.9299}}
\blacken\thicklines
\path(4609.033,1255.097)(4607.000,1115.000)(4680.998,1233.977)(4609.033,1255.097)
\thinlines
\put(2731.500,1218.929){\arc{394.717}{2.4948}{6.9299}}
\blacken\thicklines
\path(2891.033,1240.097)(2889.000,1100.000)(2962.998,1218.977)(2891.033,1240.097)
\thinlines
\put(1029.500,1233.929){\arc{394.717}{2.4948}{6.9299}}
\blacken\thicklines
\path(1189.033,1255.097)(1187.000,1115.000)(1260.998,1233.977)(1189.033,1255.097)
\thinlines
\put(1030,791){\ellipse{630}{630}}
\put(2730,791){\ellipse{630}{630}}
\put(4455,800){\ellipse{630}{630}}
\put(6166,797){\ellipse{630}{630}}
\put(7867,807){\ellipse{630}{630}}
\put(1031,792){\ellipse{540}{540}}
\put(9584,801){\ellipse{630}{630}}
\put(11295,805){\ellipse{630}{630}}
\path(2739,12)(2739,462)
\blacken\thicklines
\path(2776.500,327.000)(2739.000,462.000)(2701.500,327.000)(2776.500,327.000)
\thinlines
\path(6167,12)(6167,462)
\blacken\thicklines
\path(6204.500,327.000)(6167.000,462.000)(6129.500,327.000)(6204.500,327.000)
\thinlines
\path(1314,665)(2431,665)
\blacken\thicklines
\path(2296.000,627.500)(2431.000,665.000)(2296.000,702.500)(2296.000,627.500)
\thinlines
\path(4188,935)(3071,935)
\blacken\thicklines
\path(3206.000,972.500)(3071.000,935.000)(3206.000,897.500)(3206.000,972.500)
\thinlines
\path(3032,665)(4149,665)
\blacken\thicklines
\path(4014.000,627.500)(4149.000,665.000)(4014.000,702.500)(4014.000,627.500)
\thinlines
\path(4749,665)(5866,665)
\blacken\thicklines
\path(5731.000,627.500)(5866.000,665.000)(5731.000,702.500)(5731.000,627.500)
\thinlines
\path(5889,935)(4772,935)
\blacken\thicklines
\path(4907.000,972.500)(4772.000,935.000)(4907.000,897.500)(4907.000,972.500)
\thinlines
\path(7585,935)(6468,935)
\blacken\thicklines
\path(6603.000,972.500)(6468.000,935.000)(6603.000,897.500)(6603.000,972.500)
\thinlines
\path(6451,657)(7568,657)
\blacken\thicklines
\path(7433.000,619.500)(7568.000,657.000)(7433.000,694.500)(7433.000,619.500)
\thinlines
\path(8154,672)(9271,672)
\blacken\thicklines
\path(9136.000,634.500)(9271.000,672.000)(9136.000,709.500)(9136.000,634.500)
\thinlines
\path(9288,935)(8171,935)
\blacken\thicklines
\path(8306.000,972.500)(8171.000,935.000)(8306.000,897.500)(8306.000,972.500)
\thinlines
\path(11020,935)(9903,935)
\blacken\thicklines
\path(10038.000,972.500)(9903.000,935.000)(10038.000,897.500)(10038.000,972.500)
\thinlines
\path(9886,672)(11003,672)
\blacken\thicklines
\path(10868.000,634.500)(11003.000,672.000)(10868.000,709.500)(10868.000,634.500)
\thinlines
\path(9624,19)(9624,469)
\blacken\thicklines
\path(9661.500,334.000)(9624.000,469.000)(9586.500,334.000)(9661.500,334.000)
\thinlines
\path(2440,935)(1323,935)
\blacken\thicklines
\path(1458.000,972.500)(1323.000,935.000)(1458.000,897.500)(1458.000,972.500)
\put(953,683){\makebox(0,0)[lb]{\smash{{\SetFigFont{9}{10.8}{\rmdefault}{\mddefault}{\updefault}$1$}}}}
\put(2648,689){\makebox(0,0)[lb]{\smash{{\SetFigFont{9}{10.8}{\rmdefault}{\mddefault}{\updefault}$2$}}}}
\put(4372,690){\makebox(0,0)[lb]{\smash{{\SetFigFont{9}{10.8}{\rmdefault}{\mddefault}{\updefault}$3$}}}}
\put(6091,691){\makebox(0,0)[lb]{\smash{{\SetFigFont{9}{10.8}{\rmdefault}{\mddefault}{\updefault}$4$}}}}
\put(7794,683){\makebox(0,0)[lb]{\smash{{\SetFigFont{9}{10.8}{\rmdefault}{\mddefault}{\updefault}$5$}}}}
\put(9504,705){\makebox(0,0)[lb]{\smash{{\SetFigFont{9}{10.8}{\rmdefault}{\mddefault}{\updefault}$6$}}}}
\put(10381,1620){\makebox(0,0)[lb]{\smash{{\SetFigFont{9}{10.8}{\familydefault}{\mddefault}{\updefault}$a,c_3,c_4,c_5,c_6$}}}}
\put(11236,690){\makebox(0,0)[lb]{\smash{{\SetFigFont{9}{10.8}{\rmdefault}{\mddefault}{\updefault}$7$}}}}
\put(1966,1769){\makebox(0,0)[lb]{\smash{{\SetFigFont{9}{10.8}{\familydefault}{\mddefault}{\updefault}$c_3,c_4,c_5,c_6$}}}}
\put(3847,1566){\makebox(0,0)[lb]{\smash{{\SetFigFont{9}{10.8}{\familydefault}{\mddefault}{\updefault}$c_4,c_5,c_6$}}}}
\put(5588,1761){\makebox(0,0)[lb]{\smash{{\SetFigFont{9}{10.8}{\familydefault}{\mddefault}{\updefault}$c_3,c_5,c_6$}}}}
\put(7247,1551){\makebox(0,0)[lb]{\smash{{\SetFigFont{9}{10.8}{\familydefault}{\mddefault}{\updefault}$c_3,c_4,c_6$}}}}
\put(8926,1754){\makebox(0,0)[lb]{\smash{{\SetFigFont{9}{10.8}{\familydefault}{\mddefault}{\updefault}$c_3,c_4,c_5$}}}}
\put(15,1567){\makebox(0,0)[lb]{\smash{{\SetFigFont{9}{10.8}{\familydefault}{\mddefault}{\updefault}$b,c_3,c_4,c_5,c_6$}}}}
\put(1786,350){\makebox(0,0)[lb]{\smash{{\SetFigFont{9}{10.8}{\familydefault}{\mddefault}{\updefault}$b$}}}}
\put(3310,350){\makebox(0,0)[lb]{\smash{{\SetFigFont{9}{10.8}{\familydefault}{\mddefault}{\updefault}$b,c_3$}}}}
\put(1817,1064){\makebox(0,0)[lb]{\smash{{\SetFigFont{9}{10.8}{\familydefault}{\mddefault}{\updefault}$a$}}}}
\put(3512,1063){\makebox(0,0)[lb]{\smash{{\SetFigFont{9}{10.8}{\familydefault}{\mddefault}{\updefault}$a$}}}}
\put(5221,1049){\makebox(0,0)[lb]{\smash{{\SetFigFont{9}{10.8}{\familydefault}{\mddefault}{\updefault}$a$}}}}
\put(6909,1055){\makebox(0,0)[lb]{\smash{{\SetFigFont{9}{10.8}{\familydefault}{\mddefault}{\updefault}$a$}}}}
\put(8633,1047){\makebox(0,0)[lb]{\smash{{\SetFigFont{9}{10.8}{\familydefault}{\mddefault}{\updefault}$a$}}}}
\put(10336,1055){\makebox(0,0)[lb]{\smash{{\SetFigFont{9}{10.8}{\familydefault}{\mddefault}{\updefault}$a$}}}}
\put(5012,343){\makebox(0,0)[lb]{\smash{{\SetFigFont{9}{10.8}{\familydefault}{\mddefault}{\updefault}$b,c_4$}}}}
\put(6708,350){\makebox(0,0)[lb]{\smash{{\SetFigFont{9}{10.8}{\familydefault}{\mddefault}{\updefault}$b,c_5$}}}}
\put(8403,350){\makebox(0,0)[lb]{\smash{{\SetFigFont{9}{10.8}{\familydefault}{\mddefault}{\updefault}$b,c_6$}}}}
\thinlines
\put(9586.500,1226.929){\arc{394.717}{2.4948}{6.9299}}
\blacken\thicklines
\path(9746.033,1248.097)(9744.000,1108.000)(9817.998,1226.977)(9746.033,1248.097)
\end{picture}
}
\end{center}
\caption{NFA $\cN$ of $L^R$, $n$ odd.} 
\label{fig:reversalodd}
\end{figure}
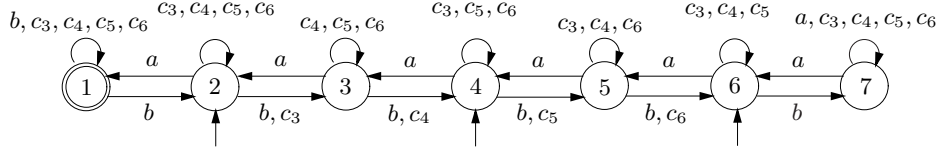

Assume initially that $n$ is odd.
Let  $S=\{s_1,\ldots,s_k\}$ be a subset of $Q$, and let $1\le s_1<\cdots< s_k\le n$.
Then  NFA $\cN$ has the following properties:
\goodbreak

\noin
{\bf P1}
If $3\le j\le n-1$, $j\in S$ and $j-1\not\in S$, then input $c_j$ deletes state $j$ from $S$ without changing any of the other states. 
\smallskip

\noin
{\bf P2}
If $3\le j\le n-1$,  $j\not\in S$,  and $j-1\in S$, then input $c_j$ adds state $j$ to $S$
without changing any of the other states.

We now examine the sets of reachable states in $\cN$.
The set $O$ of all the odd states cannot be reached. For suppose that it is reached from some set $S$. If it is reached by $a$, then $S$ must be a subset of  $E\cup\{n\}$. 
However, the successor under $a$ of such a set  $S$ also contains $n-1$ if it contains $n$.
If we use $b$, then $S$ must be a subset of $E\cup\{1\}$.
But then the successor of $S$  also contains 2 if it contains~1.
If we use $c_i$ with $i$ odd, then $S$ must be a subset of $O\setminus\{i\}$, and $S$ must also have $i-1$. But then the successor of $S$ also contains ${i-1}$, which is even, if it contains $i$.
If we use $c_i$ with $i$ even, then we also get $i$.

If $n=3$, there are no $c_i$ inputs. Set $\{2\}$  is initial, $\{1\}$ can be reached by $a$ and $\{3\}$ by $b$.
We can get $\emp$ by  $aa$,  $\{1,2\}$ and $\{2,3\}$  by $ab$ and $ba$, respectively, and $\{1,2,3\}$  by $abb$. 
Set $\{1,3\}$ is unreachable. So assume $n\ge 5$.

First, consider subsets $S$ of $M$, the set of \emph{middle states}; these are subsets of $Q$ containing neither 1 nor $n$. If $2\in S$ start with
$E=\{2,4,\ldots,n-1\}$. 
By using inputs $c_i$, delete $n-1$ or not, add $n-2$ or not, etc., until we reach 2, which cannot be removed by any $c_i$.
If $2\not\in S$, then  $S_{-(s_1-2)}$  has 2, is a subset of $M$, and so is reachable; then $S$ is reached  by  $b^{s_1-2}$ from $S_{-(s_1-2)}$.

Second, consider subsets $S$ of $Q$ containing 1 but not $n$.
If $2\in S$, start with $E$ and apply $ab$ to reach $\{1\} \cup E$. 
Each state in $E$, except  2, is without a predecessor in $\{1\} \cup E$.
Hence, by using  inputs $c_i$, we can construct any such  $S$.
If $2\not\in S$,
start with $E$ and apply $a$ to reach $O\setminus \{n\}$, where $O$ of all the odd states.
By using inputs $c_i$, we can construct any such set $S$.


Third,  examine subsets $S$ of $Q$ containing $n$ but not 1.
If $2\in S$, start with $E$ and apply $b$ to reach $E_{+1}=\{3,5,\ldots,n\}=O\setminus\{1\}$, and then apply $a$ to get $E\cup \{n\}$.
Construct any such set $S$ using inputs $c_i$.
If $2\not\in S$, then  $S$ is a subset of $\{3,\ldots,n\}$ containing $n$. 
Since the set $S_{-1}$   is a subset of $M$, it is reachable; then $S$ is reached  by  $b$ from $S_{-1}$.

Finally, consider subsets $S$ containing both 1 and $n$.
Apply $baab$ to $E$ to reach $\{1,n\}\cup E$. From this set we can reach any set containing $\{1,2,n\}$.

Now assume that $2\not\in S$.
We now show that $\{i\} \cup O$ is reachable for every even $i>2$ in $Q$.
Apply $baa$ to $E$ to reach $\{n-1\}\cup O$.
 If $i=n-1$, we are done; otherwise, delete $n-2$ and $n-1$ by $c_{n-2}$ and $c_{n-1}$ in that order. Then insert $n-3$ and $n-2$ by $c_{n-3}$ and $c_{n-2}$ in that order.
 If $i=n-3$, we are done; otherwise,  continue in this fashion.
If we reach $\{3,4,5\}$, then $i=4$, and the process stops. 

If $n=5$, then we can reach $\{1,3,4,5\}$.
From $\{1,3,4,5\}$ we can get $\{1,5\}$, $\{1,4,5\}$, and $\{1,3,4,5\}$.
We are missing only  $\{1,3,5\}$, which is unreachable.

If $n\ge7$, from $\{n-1\}\cup O$ we can reach by $c_i$ inputs all the subsets containing   $\{1,n\}$ but not $\{2\}$, except those subsets containing $n-2$ without $n-1$.
From now on,  we are interested only in the missing subsets, which are with $\{1,n\}$, without $2$, and have $n-2$ without $n-1$.
Then take $\{n-3\}\cup O$. 
From here we can reach all subsets containing   $\{1,n-2,n\}$ without $\{2,n-1\}$, except those containing $n-4$ without $n-3$. If $n=7$, then $n-4=3$, and we are missing only 
$\{1,3,5,7\}$, which is unreachable.

Continuing in this fashion, we can reach all the subsets containing 
$\{1,n\}$ but not 2, except $O$. Together with the case where $2\in S$, we have all the states containing $\{1,n\}$,  except $O$.

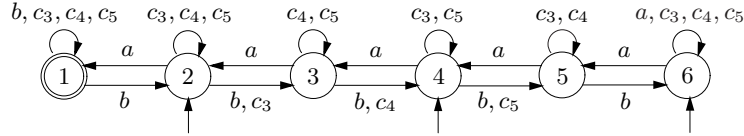
\begin{figure}[tbh]
\begin{center}
\setlength{\unitlength}{0.00037620in}
\begingroup\makeatletter\ifx\SetFigFont\undefined%
\gdef\SetFigFont#1#2#3#4#5{%
  \reset@font\fontsize{#1}{#2pt}%
  \fontfamily{#3}\fontseries{#4}\fontshape{#5}%
  \selectfont}%
\fi\endgroup%
{\renewcommand{\dashlinestretch}{30}
\begin{picture}(9614,1856)(0,-10)
\put(8581,1573){\makebox(0,0)[lb]{\smash{{\SetFigFont{9}{10.8}{\familydefault}{\mddefault}{\updefault}$a,c_3,c_4,c_5$}}}}
\put(7576.500,1233.929){\arc{394.717}{2.4948}{6.9299}}
\blacken\thicklines
\path(7736.033,1255.097)(7734.000,1115.000)(7807.998,1233.977)(7736.033,1255.097)
\thinlines
\put(5888.500,1226.929){\arc{394.717}{2.4948}{6.9299}}
\blacken\thicklines
\path(6048.033,1248.097)(6046.000,1108.000)(6119.998,1226.977)(6048.033,1248.097)
\thinlines
\put(4156.500,1233.929){\arc{394.717}{2.4948}{6.9299}}
\blacken\thicklines
\path(4316.033,1255.097)(4314.000,1115.000)(4387.998,1233.977)(4316.033,1255.097)
\thinlines
\put(2438.500,1218.929){\arc{394.717}{2.4948}{6.9299}}
\blacken\thicklines
\path(2598.033,1240.097)(2596.000,1100.000)(2669.998,1218.977)(2598.033,1240.097)
\thinlines
\put(736.500,1233.929){\arc{394.717}{2.4948}{6.9299}}
\blacken\thicklines
\path(896.033,1255.097)(894.000,1115.000)(967.998,1233.977)(896.033,1255.097)
\thinlines
\put(737,791){\ellipse{630}{630}}
\put(2437,791){\ellipse{630}{630}}
\put(4162,800){\ellipse{630}{630}}
\put(5873,797){\ellipse{630}{630}}
\put(7574,807){\ellipse{630}{630}}
\put(738,792){\ellipse{540}{540}}
\put(9291,801){\ellipse{630}{630}}
\path(2446,12)(2446,462)
\blacken\thicklines
\path(2483.500,327.000)(2446.000,462.000)(2408.500,327.000)(2483.500,327.000)
\thinlines
\path(5874,12)(5874,462)
\blacken\thicklines
\path(5911.500,327.000)(5874.000,462.000)(5836.500,327.000)(5911.500,327.000)
\thinlines
\path(1021,665)(2138,665)
\blacken\thicklines
\path(2003.000,627.500)(2138.000,665.000)(2003.000,702.500)(2003.000,627.500)
\thinlines
\path(3895,935)(2778,935)
\blacken\thicklines
\path(2913.000,972.500)(2778.000,935.000)(2913.000,897.500)(2913.000,972.500)
\thinlines
\path(2739,665)(3856,665)
\blacken\thicklines
\path(3721.000,627.500)(3856.000,665.000)(3721.000,702.500)(3721.000,627.500)
\thinlines
\path(4456,665)(5573,665)
\blacken\thicklines
\path(5438.000,627.500)(5573.000,665.000)(5438.000,702.500)(5438.000,627.500)
\thinlines
\path(5596,935)(4479,935)
\blacken\thicklines
\path(4614.000,972.500)(4479.000,935.000)(4614.000,897.500)(4614.000,972.500)
\thinlines
\path(7292,935)(6175,935)
\blacken\thicklines
\path(6310.000,972.500)(6175.000,935.000)(6310.000,897.500)(6310.000,972.500)
\thinlines
\path(6158,657)(7275,657)
\blacken\thicklines
\path(7140.000,619.500)(7275.000,657.000)(7140.000,694.500)(7140.000,619.500)
\thinlines
\path(7861,672)(8978,672)
\blacken\thicklines
\path(8843.000,634.500)(8978.000,672.000)(8843.000,709.500)(8843.000,634.500)
\thinlines
\path(8995,935)(7878,935)
\blacken\thicklines
\path(8013.000,972.500)(7878.000,935.000)(8013.000,897.500)(8013.000,972.500)
\thinlines
\path(9331,19)(9331,469)
\blacken\thicklines
\path(9368.500,334.000)(9331.000,469.000)(9293.500,334.000)(9368.500,334.000)
\thinlines
\path(2147,935)(1030,935)
\blacken\thicklines
\path(1165.000,972.500)(1030.000,935.000)(1165.000,897.500)(1165.000,972.500)
\put(660,683){\makebox(0,0)[lb]{\smash{{\SetFigFont{9}{10.8}{\rmdefault}{\mddefault}{\updefault}$1$}}}}
\put(2355,689){\makebox(0,0)[lb]{\smash{{\SetFigFont{9}{10.8}{\rmdefault}{\mddefault}{\updefault}$2$}}}}
\put(4079,690){\makebox(0,0)[lb]{\smash{{\SetFigFont{9}{10.8}{\rmdefault}{\mddefault}{\updefault}$3$}}}}
\put(5798,691){\makebox(0,0)[lb]{\smash{{\SetFigFont{9}{10.8}{\rmdefault}{\mddefault}{\updefault}$4$}}}}
\put(7501,683){\makebox(0,0)[lb]{\smash{{\SetFigFont{9}{10.8}{\rmdefault}{\mddefault}{\updefault}$5$}}}}
\put(9211,705){\makebox(0,0)[lb]{\smash{{\SetFigFont{9}{10.8}{\rmdefault}{\mddefault}{\updefault}$6$}}}}
\put(1493,350){\makebox(0,0)[lb]{\smash{{\SetFigFont{9}{10.8}{\familydefault}{\mddefault}{\updefault}$b$}}}}
\put(3017,350){\makebox(0,0)[lb]{\smash{{\SetFigFont{9}{10.8}{\familydefault}{\mddefault}{\updefault}$b,c_3$}}}}
\put(1524,1064){\makebox(0,0)[lb]{\smash{{\SetFigFont{9}{10.8}{\familydefault}{\mddefault}{\updefault}$a$}}}}
\put(3219,1063){\makebox(0,0)[lb]{\smash{{\SetFigFont{9}{10.8}{\familydefault}{\mddefault}{\updefault}$a$}}}}
\put(4928,1049){\makebox(0,0)[lb]{\smash{{\SetFigFont{9}{10.8}{\familydefault}{\mddefault}{\updefault}$a$}}}}
\put(6616,1055){\makebox(0,0)[lb]{\smash{{\SetFigFont{9}{10.8}{\familydefault}{\mddefault}{\updefault}$a$}}}}
\put(8340,1047){\makebox(0,0)[lb]{\smash{{\SetFigFont{9}{10.8}{\familydefault}{\mddefault}{\updefault}$a$}}}}
\put(4719,343){\makebox(0,0)[lb]{\smash{{\SetFigFont{9}{10.8}{\familydefault}{\mddefault}{\updefault}$b,c_4$}}}}
\put(6415,350){\makebox(0,0)[lb]{\smash{{\SetFigFont{9}{10.8}{\familydefault}{\mddefault}{\updefault}$b,c_5$}}}}
\put(8373,335){\makebox(0,0)[lb]{\smash{{\SetFigFont{9}{10.8}{\familydefault}{\mddefault}{\updefault}$b$}}}}
\put(1875,1574){\makebox(0,0)[lb]{\smash{{\SetFigFont{9}{10.8}{\familydefault}{\mddefault}{\updefault}$c_3,c_4,c_5$}}}}
\put(5497,1573){\makebox(0,0)[lb]{\smash{{\SetFigFont{9}{10.8}{\familydefault}{\mddefault}{\updefault}$c_3,c_5$}}}}
\put(7209,1567){\makebox(0,0)[lb]{\smash{{\SetFigFont{9}{10.8}{\familydefault}{\mddefault}{\updefault}$c_3,c_4$}}}}
\put(3795,1574){\makebox(0,0)[lb]{\smash{{\SetFigFont{9}{10.8}{\familydefault}{\mddefault}{\updefault}$c_4,c_5$}}}}
\put(15,1574){\makebox(0,0)[lb]{\smash{{\SetFigFont{9}{10.8}{\familydefault}{\mddefault}{\updefault}$b,c_3,c_4,c_5$}}}}
\thinlines
\put(9293.500,1226.929){\arc{394.717}{2.4948}{6.9299}}
\blacken\thicklines
\path(9453.033,1248.097)(9451.000,1108.000)(9524.998,1226.977)(9453.033,1248.097)
\end{picture}
}
\end{center}
\caption{NFA $\cN$ of $L^R$, $n$ even.} 
\label{fig:reversaleven}
\end{figure}

The case where $n$ is even is similar. The NFA $\cN$ is shown in Fig.~\ref{fig:reversaleven} for $n=6$.
By an argument similar to that for $n$ odd,  $O$ cannot be reached. 

Any subset of $M=Q\setminus \{1,n\}$ can be reached as follows.
If $2\in S$, apply $b$ to $E$ to get $O\setminus\{1\}$, and then $a$ to get to $E\setminus\{n\}$.
Now any subset of $M$ containing $2$ can be reached by inputs  $c_i$.
If $2\not\in S$, then any subset of $M\setminus \{2\}$ can be reached from $O\setminus\{1\}$ by  inputs $c_i$.

Second, consider subsets $S$ of $Q$ containing 1 but not $n$.
If $2\in S$, start with $E$ and apply $ba$ to reach $E\setminus \{n\}$. 
Then apply $ab$ to get $E\setminus \{n\}\cup \{1\}$.
Now any subset of $\{1\}\cup M$ containing $\{1,2\}$ can be reached by inputs  $c_i$.
If $2\not\in S$, start with $E$ and apply $baa$ to reach $O\setminus \{n-1\}$.
By using inputs $c_i$, we can construct any subset $S$ of $\{1\}\cup M$ containing 1 and not 2,  except the subsets that have $\{n-3,n-1\}$ without $n-2$.
In case $n=4$, we can reach $\{1,2\}$, $\{1,2,3\}$, and $\{1\}$, but not $\{1,3\}$.
From now on,  we are interested only in the missing subsets.
As in the even case, we can get subsets containing $\{n-3,n-1\}$ without $n-2$ by deleting
$n-3$ and $n-2$,  adding $n-4$, and re-inserting $n-3$. Now we are unable to reach
states having $\{n-5,n-3\}$ without $n-4$. 
We verify that $\{i\} \cup O$ is reachable for every even $i$ with $4\le i\le i-2$, and continue as  in the odd case.
We can keep moving  this problem to the left, until we reach $\{3,4,5\}$. Then state 4 cannot be removed because $O$ is not reachable.

Third,  examine subsets $S$ of $Q$ containing $n$ but not 1.
If $2\in S$, all such subsets are reachable by inputs $c_i$ from $E$.
If $2\not\in S$, then  $S$ is a subset of $\{3,\ldots,n\}$ containing $n$. 
Since  $S_{-1}$   is a subset of $M$, it is reachable; then $S$ is reached  by  $b$ from $S_{-1}$.

Finally, consider subsets $S$ containing both 1 and $n$.
If $2\in S$, apply $ab$ to reach $\{1\}\cup E$. From here we can reach any set containing $\{1,2,n\}$ by inputs $c_i$.
If $2\not\in S$, we reach $O\cup \{n\}$ from $E$ by $a$.
From here we can reach any set containing $\{1,n\}$ but not 2 by inputs $c_i$.

We still need to verify that all the reachable subsets are pairwise distinguishable. 
State $i$, and only state $i$, accepts $a^{i-1}$.  Hence, if $S,T\subseteq Q$ and $S$ and $T$ differ by state $i$, then they are distinguishable by $a^{i-1}$.
\qed
\end{proof}
\section{Unary Languages}
\label{sec:unary}
The case of unary languages is special. For regular unary languages, the tight bounds for each boolean operation $K\circ L$, product $KL$, star $L^*$, and reversal $L^R$ are $mn$, $mn$, $n^2-2n+2$, and $n$, respectively~\cite{YZS94}. With the exception of the bound for reversal, these bounds cannot be met by star-free unary languages.

\begin{theorem}
\label{thm:unary}
Let $K$ and $L$ be unary star-free languages with quotient complexities $m$ and $n$, respectively. \\
  \hglue10pt  1. 
For each boolean operation  $\circ$, $\kappa(K\circ L)\le {\rm max}(m,n)$ and the bound is tight.\\
 \hglue10pt  2. 
For product,  $\kappa(KL)\le m+n-1$, 
and the bound is tight.\\ 
 \hglue10pt  3. For the star, the tight bound is
 \begin{align*}
	\kappa(L^*) \le \left\{\begin{array}{cl}
		2,	& \text{ if }	 n=1; \\
    		n,	& \text{ if }	 2\le n\le 5;\\
		n^2-7n+13,	& \text{ otherwise }.
         	\end{array}\right.
  \end{align*}\\
 \hglue10pt  4. For reversal, $\kappa(L^R)=n$.
\end{theorem}

\begin{proof} If a unary star-free language $L$ is finite and $\kappa(L)=n$, its longest word has length $n-2$; if it is infinite,   the longest word not in $L$ has length  $n-2$.\\
 \hglue10pt  1. 
One verifies that $\kappa(K\circ L)\le {\rm max}(m,n)$. 
The witness languages are  $K=a^{m-2}$ and $L=a^{n-2}$  for union and symmetric difference, 
$K'=a^{m-1}a^*$ and $L'=a^{n-1}a^*$ for intersection, and $K'$ and $\ol{L'}$ for difference, since $K'\setminus \ol{L'}=K'\cap L'$.
 \hglue10pt  2. 
One verifies that $\kappa(KL)\le m+n-1$, and   $K=a^{m-1}a^*$ and $L=a^{n-1}a^*$ are witnesses.\\
 \hglue10pt  3. 
 If $L$ is infinite, then $L\supseteq a^{n-1}a^*$, and $L^*\supseteq a^{n-1}a^*$; hence
$\kappa(L^*)\le n$.
For $n=1,2,3,4,5$, the bounds actually met in the infinite case are 1, 1, 3, 4, 5, respectively.
If $L$ is finite,  it must contain $a^{n-2}$, and if it has $a$, then $\kappa(L^*)=1$.
The tight bounds for finite unary star-free languages are
2, 2, 1, 2, 3, respectively. 
Hence the tight bounds for all unary star-free languages for the first five values of $n$ are 2, 2, 3, 4, 5, and the witnesses are $\emp$, $\eps$, $a^2a^*$, $a^3a^*$, and $a^4a^*$, respectively.

It was shown in~\cite{CCSY01} that for a finite unary language $L$, $\kappa(L^*)\le n^2-7n+13$ for $n\ge 5$. 
For $n>6$, this bound applies here, and
a witness is $a^{n-3}\cup a^{n-2}$.\\
 \hglue10pt  4. 
For unary languages, we have $L^R=L$; hence $\kappa(L^R)=\kappa(L)$.
\qed
\end{proof}

\section{Conclusions}
\label{sec:conclusions}

We have shown that all the commonly used regular operations in the class of star-free languages meet the quotient complexity bounds of arbitrary regular languages. The only exceptions are in the product for $n=2$, reversal, and operations on unary languages.


\providecommand{\noopsort}[1]{}

\end{document}